\documentclass[a4paper,cite]{article}

\usepackage{latexsym,epsfig}

\usepackage[latin1]{inputenc}

\usepackage{graphicx}
\usepackage{color,pst-plot}

\usepackage{amsmath}
\usepackage {amsfonts,amssymb,amstext}
\usepackage{psfrag}

\newcommand{\beq}[1]{
\begin{equation}\label{#1}}
\newcommand{\eeq}{\end{equation}}
\newcommand{\bea}[1]{
\marginpar{\small\textsf{#1}}
\begin{eqnarray}\label{#1}}
\newcommand{\eea}{\end{eqnarray}}
\def\noi{\noindent}

\def\bea{\begin{eqnarray}}
\def\beqa{\begin{eqnarray}}
\def\eea{\end{eqnarray}}
\def\eqa{\end{eqnarray}}
\def\beas{\begin{eqnarray*}}
\def\eeas{\end{eqnarray*}}
\def\beqas{\begin{eqnarray*}}
\def\eqas{\end{eqnarray*}}
\def\beq{\begin{equation}}
\def\eeq{\end{equation}}
\def\beqd{\begin{displaymath}}
\def\eeqd{\end{displaymath}}
\def\eqd{\end{displaymath}}

\def\beeq{\begin{eqnarray}} \def\eeeq{\end{eqnarray}}

\newcommand{\eq}{\end{equation}}

\newcommand{\zb}{\bar{z}}

\newcommand{\be}{\begin{equation}}
\newcommand{\ee}{\end{equation}}{

\DeclareMathAlphabet{\eusm}{U}{}{}{}
\SetMathAlphabet\eusm{normal}{U}{eus}{m}{n}
\SetMathAlphabet\eusm{bold}{U}{eus}{b}{n}
\DeclareMathAlphabet{\mathpzc}{OT1}{pzc}{m}{it}

\input epsf

\begin{document}

\begin{titlepage}

\begin{flushright}
CPHT-RR006.0308\\
LPT-08-33\\
arXiv:yymm.nnnn\\

\end{flushright}

\vspace*{0.2cm}
\begin{center}
{\Large {\bf Diphoton
Generalized Distribution Amplitudes
}}\\[2 cm]

{\bf M. El Beiyad}~$^{a, b}$, {\bf B. Pire}~$^a$, {\bf  L. Szymanowski}~$^{a,b,c}$ {\bf and S. Wallon}~$^{ b}$\\[1cm]

$^a$  {\it Centre  de Physique Th{\'e}orique, \'Ecole Polytechnique, CNRS,
   91128 Palaiseau, France}\\[0.5cm]
$^b$ {\it LPT, Universit\'e d'Orsay, CNRS, 91404 Orsay, France}\\[0.5cm]
$^c$ {\it Soltan Institute for Nuclear Studies, Warsaw, Poland}
\end{center}

\vspace*{3.0cm}

\begin{abstract}
We calculate the leading order diphoton generalized distribution amplitudes by calculating the amplitude
of the process $\gamma^* \gamma \to \gamma \gamma$ in the low energy and high photon virtuality region
 at the Born order and in the leading logarithmic approximation. As in the case of the anomalous photon structure
 functions, the $\gamma \gamma$ generalized distribution amplitudes exhibit a characteristic  $\ln Q^2$
 behaviour and obey inhomogeneous QCD evolution equations.
  \end{abstract}
\vspace{1cm}

\end{titlepage}

\pagebreak
\section{\normalsize Introduction}
The photon is a much interesting object for QCD studies. Its pointlike coupling to quarks enables to calculate perturbatively part of its wave function. In the case of a virtual photon, this perturbative part is leading at large virtuality; a twist expansion generates non-leading components of the photon distribution amplitude~\cite{Braun}, from which the lowest order one is chiral-odd and proportional to the magnetic susceptibility of the vacuum. The study of the two photon state is kinematically richer and is thus a most welcome theoretical laboratory for the study of exclusive hard reactions. 

The parton content of the photon has been the subject of many studies since the seminal paper by Witten~\cite{Witten}. In a recent paper~\cite{FPS}  we extended the notion of anomalous parton distribution in a photon   to the case of generalized parton distributions (GPDs) used for the factorized description of the non-diagonal kinematics of deeply virtual Compton scattering (DVCS), $\gamma^*(q) \gamma \to \gamma \gamma$, namely at large energy and small hadronic momentum transfer but large photon virtuality ($Q^2=-q^2$). 

The two meson GDAs generalize \cite{GDA,GDArho} the concept of usual distribution amplitude of mesons and baryons introduced long time ago in the studies of exclusive processes \cite{BL}. They describe the coupling of a quark-antiquark (or gluon-gluon) pair to a pair of  mesons, and are related by crossing to the meson GPDs. In the same way, the  two photon generalized distribution amplitudes (GDAs)  which describes the coupling of a quark antiquark (or gluon-gluon) pair to a pair of photons, are related by crossing to the photon GPDs.

We thus study here the scattering amplitude of the $\gamma^*(q) \gamma \to \gamma \gamma$ process in the near threshold kinematics, namely at small $s$  and large $-t \sim Q^2$, at large $Q^2$ in the leading order of the electromagnetic coupling. This enables us to define and calculate perturbatively the Born approximation of the  diphoton GDAs.
This Born order contribution enters the QCD evolution equations as an inhomogeneous
term and we exhibit the properties of the evolved solution.

On the phenomenlogical side, we do not expect any  study of this process to be feasible since the production of hadronic states that decay into a pair of photons - mostly pseudoscalar neutral mesons - will strongly dominate the counting rates of this purely electromagnetic process.

The analysis presented in this paper is reminiscent of the
perturbative calculation of the two $\rho$ GDAs in terms of the single $\rho$
DAs \cite{TDA-GDA}, for which both incoming photon were chosen to be
hard in order to justify this factorization of the GDAs.

\section{\normalsize The  $\gamma^*(q) \gamma(p_1) \to \gamma(q') \gamma(p_2)$ process
in the threshold region}

\begin{figure}[h]
$\begin{array}{ccc}
\includegraphics[width=5cm]{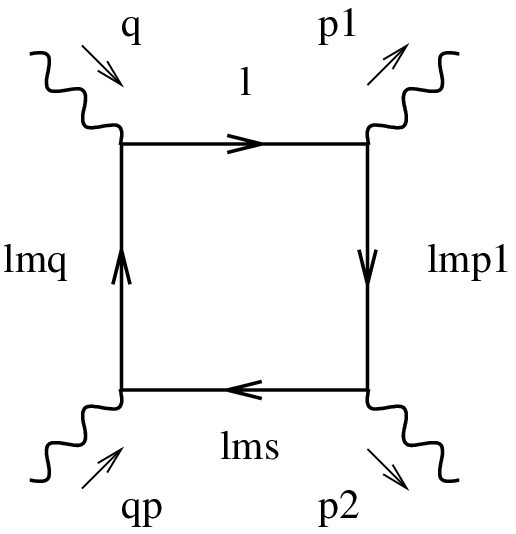}&
\includegraphics[width=4.2cm]{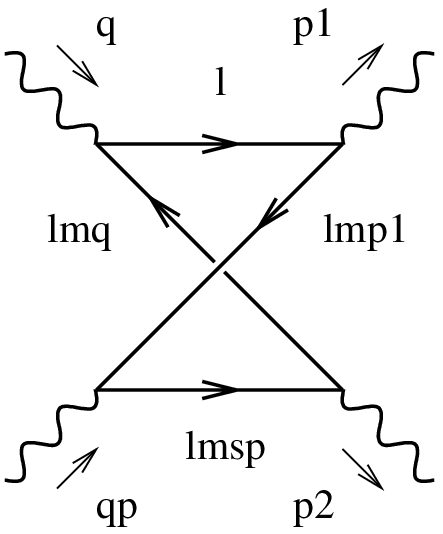}&
\includegraphics[width=5.1cm]{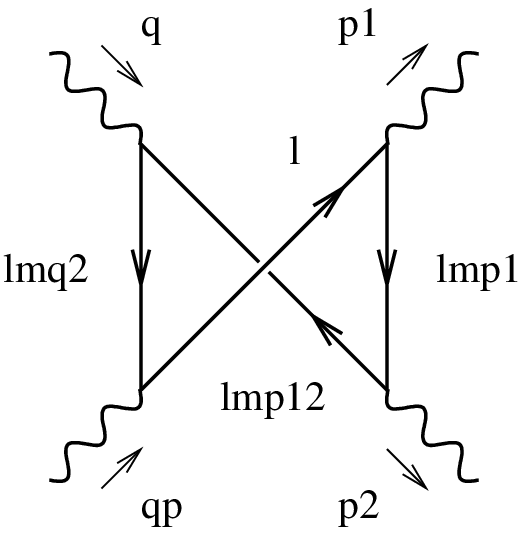} \\
\\
A & B & C\\
\\
\includegraphics[width=5cm]{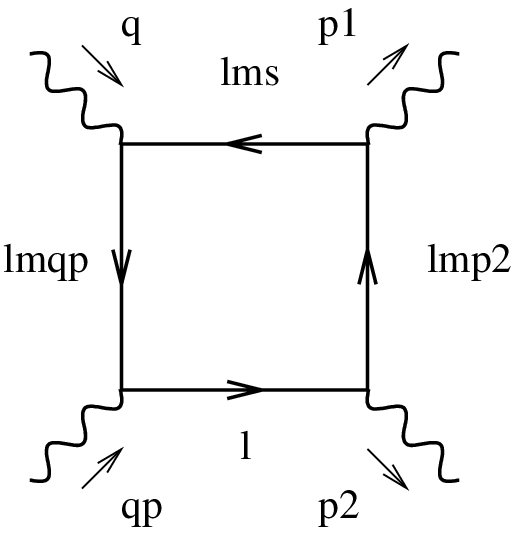}&
\includegraphics[width=4.2cm]{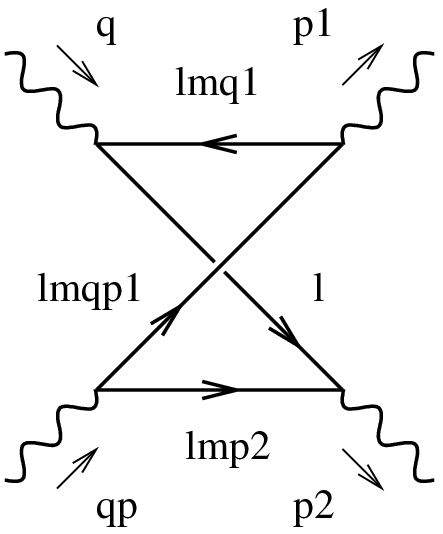}&
\includegraphics[width=5.1cm]{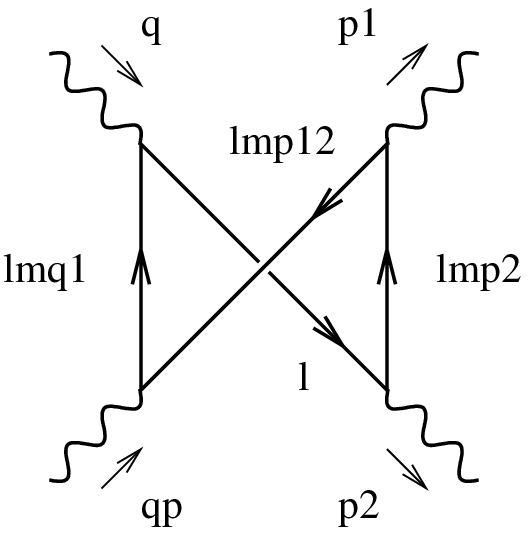} \\
\\
\bar{A} & \bar{B} & \bar{C}\\
\end{array}$
\caption[]{\small The Born order diagrams for $\gamma^* \;\gamma \to \gamma  \; \gamma$}
\label{FigDiagrams}
\end{figure}

 Two photon production in  Compton scattering  on  a photon target
\begin{equation}
\gamma^*(q) \gamma(q') \to \gamma(p_{1}) \gamma(p_2)
\label{dvcs}
\end{equation}
involves, at leading order in $\alpha_{em}$, and zeroth order in  $\alpha_{S}$ the six Feynman diagrams 
of Fig.\ref{FigDiagrams} with quarks in the loop. Note that we do not consider in this paper the pure QED content of this amplitude, involving a lepton in the loop.  It can  straightforwardly be obtained from the results below.
\psfrag{q}[cc][cc]{$\! q$}
\psfrag{qp}[cc][cc]{$\! \! q'$}
\psfrag{p1}[cc][cc]{$\ \ p_1$}
\psfrag{p2}[cc][cc]{$\ \ p_2$}
\psfrag{l}[cc][cc]{$l$}
\psfrag{lmq}[cc][cc]{$l-q$}
\psfrag{lms}[cc][cc]{$l-p_1-p_2$}
\psfrag{lmsp}[cc][cc]{$l-q_1+p_2$}
\psfrag{lmp1}[cc][cc]{$l-p_1$}
\psfrag{lmq2}[cc][cc]{$l-q'$}
\psfrag{lmq1}[cc][cc]{$l-q$}
\psfrag{lmqp1}[cc][cc]{$l-q+p_1$}
\psfrag{lmqp}[cc][cc]{$l-q'$}
\psfrag{lmp2}[cc][cc]{$l-p_2$}
\psfrag{lmp12}[cc][cc]{$l-p_1-p_2$}

For simplicity, we restrict in this paper to the threshold kinematics where $W^2 = (p_{1}+p_{2})^2 = 0$.
This simplifies greatly the tensorial structure of the amplitude  while still preserving the richness
of the  skewedness ($\zeta$) dependence of GDAs, but forbids any impact parameter 
interpretation of the GDAs~\cite{GDArho}. Our conventions for the kinematics are the following:

\begin{equation}\nonumber
q= p - \frac{Q^2}{s}n\ , ~~~~~~~~~~ q'= \frac{Q^2}{s}n\ ,
\end{equation}

\begin{equation}\nonumber
p_1=\zeta p\ , ~~~~~~~~~~ p_2= \bar \zeta p\ , ~~~~~~~~~~   \bar \zeta = 1-\zeta ,
\end{equation}
where $p$ and $n$ are two light-cone Sudakov vectors and $2 p\cdot n =s $.
The momentum $l$ in the quark loop is parametrized as
\begin{equation}
l^\mu = z p^\mu +\beta n^\mu + l_T\ ,
\label{Sudakov}
\end{equation}
with $l_T^2=-{\mathbf l^2}$.
The process involves one virtual and 
three real photons and its amplitude can 
be written as
\begin{equation}
A = \epsilon_\mu\epsilon'_\nu{\epsilon_1}^*_\alpha{\epsilon^*_2}_\beta T^{\mu\nu\alpha\beta},
\end{equation}
where in our (forward) kinematics  the four
photon polarization vectors
$\epsilon(q)$, $\epsilon'(q')$, $\epsilon_1(p_1)$ and $\epsilon_2(p_2)$ are 
transverse with respect to the Sudakov vectors $p$ and $n.$

The tensorial decomposition of $T^{\mu\nu\alpha\beta}$ reads \cite{BGMS75}
\begin{equation}
T^{\mu\nu\alpha\beta} (W=0) = \frac{1}{4}g^{\mu\nu}_Tg^{\alpha\beta}_T W_1+
\frac{1}{8}\left(g^{\mu\alpha}_Tg^{\nu\beta}_T 
+g^{\nu\alpha}_Tg^{\mu\beta}_T -g^{\mu\nu}_Tg^{\alpha\beta}_T \right)W_2
+ \frac{1}{4}\left(g^{\mu\alpha}_Tg^{\nu\beta}_T - g^{\mu\beta}_Tg^{\alpha\nu}_T\right)W_3\, ,
\end{equation}
and it involves three scalar functions $W_i$, $i=1,2,3$.

The integration over $l$ is performed as usual within the Sudakov representation, using
$$d^4l = \frac{s}{2} \, dz \, d\beta \, d^2 l_{T} \rightarrow \frac{\pi s}{2} \, dz \, d\beta \, d {\mathbf l^2}\,.$$
Before starting the explicit computation, we note that in order to interpret our result in terms of  factorized quantities, we will keep our expressions unintegrated with respect to $z.$  In all  the 
following calculations, the mass
of the quark will play the role of an infrared regulator.
Let us briefly present  our calculations for diagram A, and then give final results for the
other diagrams. We get for $ W_{1}^{A}$
\begin{equation}
W_{1}^{A}=  -i \frac{s e_{q}^4 N_{C}}{32 \pi^3} \int
  \frac{dz\; d\beta\; d{\mathbf l^2}\;\;\;Tr A_{1}}{[(l-q)^2 -m^2+i\eta] [(l-p_{1})^2-m^2+i\eta]
 (l^2-m^2+i\eta)[(l-q-q')^2-m^2+i\eta]}\,,
\label{ReW}
\end{equation}
where $Tr A_1 = \mathcal{T}r [\gamma_{T}^\mu (\hat l - \hat q + m) \gamma_{T}^\mu(\hat l-\hat p_{1} - \hat p_{2}+m)
\gamma_{T}^\alpha (\hat l -\hat p_{1}+m)\gamma_{T}^\alpha (\hat l +m)]$ and $\mathcal{T}r$ means the trace over spinorial indices.

We first integrate over  $\beta$ using the Cauchy theorem. The propagators induce poles in the complex $\beta$-plane
with values
\begin{eqnarray}
&& \beta_{1} =   \frac{{\bold l^2} + m^2 -i\eta}{z s}\,, \hspace{2cm}\beta_{2}=
\frac {{\bold l^2} + m^2 -i\eta}{(z - \zeta) s} \,,\nonumber \\
&&\beta_{3}=  - \frac{{\bold l^2} + m^2 + \bar z Q^2 -i\eta}{\bar z s}\,, \qquad \beta_{4}= -
\frac{{\bold l^2} + m^2 -i\eta}{\bar z s} \;.
\label{defbeta}
\end{eqnarray}
Since the four poles lie all below the real axis for $z>1$ and lie all above the real
axis for $z<0$, the only region 
where the amplitude may not vanish is   $1 > z > 0$.
One identifies two different regions:
\begin{itemize}
\item  the    region where  $\zeta<z<1$, for which one may close the contour 
in the lower half plane and get the contribution of the poles $\beta_{1}$ and $\beta_2$\,,
\item   the  region where  $0<z<\zeta$, for which one may close the contour 
in the lower half plane and take the contribution of the pole $\beta_{1}$\,.
\end{itemize}
This is reminiscent of the different  regions encountered in the kinematics of the generalized parton distributions $H(x,\xi,t)$, with the boundaries controlled by the relative values of $x$ and $\xi$, where $x$ and $\xi$ are analogous to our $z$ and $\zeta$ variables.\\
The result of $\beta$ integration takes  thus the form
\begin{equation}
I^{A_1} =  -2i\pi\int_{\zeta}^1 dz\, \int\, d{\bold l^2} \left(\frac{Tr A _1(\beta = \beta_{1})}{DA_{\beta_1}} +\frac{Tr A _1(\beta = \beta_{2})}{DA_{\beta_2}}\right)-2i\pi\int_{0}^\zeta dz\, \int\, d{\bold l^2} \frac{Tr A_1 (\beta = \beta_{1})}{DA_{\beta_1}},
\label{Idiv}
\end{equation}
where $DA_{\beta_{i}}$ denote the value of the product of propagators at the pole $\beta_i$.\\
A number of technical simplifications are now helpful. Firstly, one easily verifies that, at the leading logarithmic approximation we are interested in, the trace $Tr A_{1}$ may be simplified by taking the limit $m^2\to 0$.
Secondly,  each integral over ${\bold l^2}$ in Eq.(\ref{Idiv}) is UV divergent. However, it is a well-known classical result of QED  that the sum of integrals 
corresponding to the six diagrams of Fig.\ref{FigDiagrams} is UV finite, so we separate UV divergent terms of each diagram in an algebraic way, and we show that the UV finiteness appears for the sum of diagrams A, B and C and for the sum of diagrams D, E and F separately.

 The traces are simple polynomials in ${\bold l^2}$, which may be written as 
 $Tr A_{1} (\beta = \beta_{i})\sim \alpha_{i}{\bold l^4} + \gamma_{i}Q^2 {\bold l^2}+ \delta_{i} Q^4\,,$ where $\alpha_i$, $\gamma_i$ and $\delta_i$ are dimensionless functions of $z$ and $\zeta$.
Power counting in ${\bold l^2}$ shows that the $\alpha_{i}{\bold l^4}$ term in these integrals is ultraviolet divergent, since $DA_{\beta_i}$ behaves as ${\bold l^6}.$ 
Our aim is to recover the UV finiteness of the sum of diagrams 
before performing the integration over $z$.  This cancellation of UV divergences occurs separately in each interval $z\in[0,\zeta]$ and $z\in[\zeta,1]$. Let us concentrate for definiteness on the interval $[0,\zeta]$.
The UV divergent part of $I^{A_1}$ in this interval  has the form :
\begin{equation}
-2i\pi\int_0^\zeta dz\int\frac{d{\bold l^2}}{l^2}a_1^{div}(z,\zeta),
\label{div}
\end{equation}
where $a_1^{div}$ is a definite function of $z$ and $\zeta$. The contributions from diagrams B and C lead to similar equations as Eq.(\ref{div}) with $a_1^{div}$ replaced respectively by $b_1^{div}$ and $c_1^{div}$. The cancellation of UV divergences in the sum of A, B and C diagrams is manifest through the fact that $a_1^{div}+b_1^{div}+c_1^{div}=0$.
This relation allows us to extract the convergent part of the second term of Eq.(\ref{Idiv}) by adding and subtracting a term which reproduces the same UV divergence but which is IR finite:
\begin{eqnarray}
I^{A_1}_{div} &=&-2i\pi\int_{0}^\zeta dz\, \int\, d{\bold l^2}\left( \frac{Tr A_1 (\beta = \beta_{1})}{DA_{\beta 1}}-\frac{\bold l^4}{(\bold l^2+m^2)^3}a_1^{div}(z,\zeta)\right) \nonumber \\
&-&2i\pi\int_0^\zeta dz\int d\bold l^2\frac{\bold l^4}{(\bold l^2+m^2)^3}a_1^{div}(z,\zeta)
\label{div2}
\end{eqnarray}
 The first term in Eq.(\ref{div2}) is, by construction, both UV and IR finite
and can be computed, whereas  the second term  in Eq.(\ref{div2}) cancels out in the sum of diagrams A, B and C and thus does not need to be evaluated. The same procedure is applied to the contribution coming from the interval $z\in[\zeta,1]$. 
Moreover, the contributions from the diagrams D, E and F are considered in the same way.

The final result of our calculation of the  amplitude $W_1$ reads, in the leading log $Q^2$ approximation
\begin{eqnarray}
 W_{1}  &=& \frac{e_q^4N_C}{2\pi^2}\int_0^1dz\ (2z-1)\left[\frac{2z-\zeta}{z\bar{\zeta}}\theta(z-\zeta)+\frac{2z-1-\zeta}{\bar{z}\zeta}\theta(\zeta-z) \right. \nonumber \\
 &+&\left. \frac{2z-\bar{\zeta}}{z\zeta}\theta(z-\bar{\zeta})+\frac{2z-1-\bar{\zeta}}{\bar{z}\bar{\zeta}}\theta(\bar{\zeta}-z)\right]\log\frac{m^2}{Q^2}\,.
 \label{W1}
\end{eqnarray}
The amplitudes $W_{2}$ and $W_{3}$  are calculated in the same way and we get
\begin{equation}
 W_{2} =0
\label{W2}
\end{equation}
and
\begin{eqnarray}
 W_{3}  &=& -\frac{e_q^4N_C}{2\pi^2}\int_0^1dz\ \left[\frac{\zeta}{z\bar{\zeta}}\theta(z-\zeta)-\frac{\bar{\zeta}}{\bar{z}\zeta}\theta(\zeta-z) \right. \nonumber \\
 &-&\left. \frac{\bar{\zeta}}{z\zeta}\theta(z-\bar{\zeta})+\frac{\zeta}{\bar{z}\bar{\zeta}}\theta(\bar{\zeta}-z)\right]\log\frac{m^2}{Q^2}.
 \label{W3}
\end{eqnarray}

In the following section we will interpret the results (\ref{W1}) and (\ref{W3}) from the point of view of QCD factorization based on the operator product
 expansion, yet still in the zeroth order of the QCD coupling constant and in the leading logarithmic approximation. The crucial point is to note that the final contribution to this amplitude involves mixing of operators constructed from quark fields with operators constructed from photon fields \cite{Witten}. This mixing can be understood by denoting the integrands of Eq.(\ref{W1}) and Eq.(\ref{W3}) by ${\cal F}(z,\zeta)\log\frac{m^2}{Q^2}$ and rewriting them by using the obvious identity
 \begin{equation}
{\cal F}(z,\zeta) \; \log \frac{m^2}{Q^2} = {\cal F}(z,\zeta) \;\log \frac{m^2}{M_{F}^2} + {\cal F}(z,\zeta) \;\log \frac{M_{F}^2}{Q^2}\;,
\label{stupid}
\end{equation}
where $M_{F}$ corresponds to an arbitrary QCD factorization scale. As will be shown below the first term with $\log \frac{m^2}{M_{F}^2}$
may be identified with the quark GDA of the photon, 
whereas the second term with $\log \frac{M_{F}^2}{Q^2}$
corresponds to the so-called photon GDA of the photon, coming from the  matrix element of
the correlator field from photonic fields which contributes at the same order in
$\alpha_{em}$ as the quark correlator to the scattering amplitude.
The choice $M_{F}^2= Q^2$ will allow to express the  amplitude only 
in terms of the quark-antiquark fragmentation into two  photons.

\section{\normalsize  QCD factorization of the DVCS amplitude on the photon}

We first consider 
two quark non local correlators on the light cone and their matrix elements between the vacuum and a diphoton state which define the diphoton GDAs $\Phi_1$, $\Phi_3\,$:

\begin{equation}
\label{Fqa}
\hspace{-1cm}F^q = \int \frac{dy}{2\pi} e^{i(2z-1)\frac{y}{2}}\langle \gamma(p_{1})  \gamma(p_{2})| \bar q(-\frac{y}{2}N)
 \gamma.N q(\frac{y}{2}N)|0 \rangle = \frac{1}{2}g_\bot^{\mu\nu}\epsilon_\mu^*(p_1)\,\epsilon_\nu^*(p_2)\,\Phi_1(z,\zeta,0)
\end{equation}
and
\begin{equation}
\label{Fqta}
\tilde F^q = \int \frac{dy}{2\pi} e^{i(2z-1)\frac{y}{2}}\langle\gamma(p_{1})  \gamma(p_{2})| \bar q(-\frac{y}{2}N)
 \gamma.N \gamma^5 q(\frac{y}{2}N)|0 \rangle= -\frac{i}{2}\epsilon^{\mu\nu pN}\epsilon_\mu^*(p_1)\,\epsilon_\nu^*(p_2)\,\Phi_3(z,\zeta,0)\,,
\end{equation}
where we note $N= n/n.p$, $\epsilon^{\mu\nu pN} = \epsilon^{\mu\nu\alpha\beta}p_\alpha N_\beta$, with $\epsilon^{0123}=1$, and where we did not write explicitely, for simplicity of notation, neither
the electromagnetic nor the gluonic Wilson lines. We will also define the matrix elements of photonic correlators
\begin{equation}
\label{Fpa}
F^\gamma = \int \frac{dy}{2\pi} e^{i(2z-1)\frac{y}{2}}\langle \gamma(p_{1})  \gamma(p_{2})|F^{N\mu}(-\frac{y}{2}N)F_\mu^N(\frac{y}{2}N) |0 \rangle 
\end{equation}
and
\begin{equation}
\label{Fpta}
\tilde F^\gamma = \int \frac{dy}{2\pi} e^{i(2z-1)\frac{y}{2}}\langle\gamma(p_{1})  \gamma(p_{2})|F^{N\mu}(-\frac{y}{2}N)\tilde F_\mu^N(\frac{y}{2}N) |0 \rangle,
\end{equation}
where $F^{N \mu} = N_\nu F^{\nu \mu}$ and $\tilde F^{\mu\nu}=\frac{1}{2}\epsilon^{\mu\nu\rho\sigma}F_{\rho\sigma}$, which mix  with correlators (\ref{Fqa}) and (\ref{Fqta}), but 
contrarily to the quark correlator matrix element, are non zero at order $\alpha_{em}^0$ \cite{Witten}.

The quark correlator matrix elements, calculated in the lowest order of $\alpha_{em}$ and $\alpha_{S}$, 
suffer from ultraviolet divergences, which we regulate 
through the usual dimensional regularization procedure, with  $d= 4+2\epsilon$. 
We obtain (with  $\frac{1}{\hat\epsilon} = \frac{1}{\epsilon} +\gamma_{E}-\log 4\pi$)
\begin{equation}
F^q= -\frac{N_C\,e_{q}^2}{4\pi^2} g_T^{\mu\nu}\epsilon^*_\mu (p_{1})\epsilon^*_\nu(p_{2}) 
 \left[\frac{1}{\hat\epsilon} + \log{m^2}\right] F(z,\zeta)\,,
\label{Fgam}
\end{equation}
with 
\begin{equation}
\label{Fz}
 F(z,\zeta) = \frac{\bar{z}(2z-\zeta)}{\bar{\zeta}}\theta(z-\zeta)+\frac{\bar{z}(2z-\bar{\zeta})}{\zeta}\theta(z-\bar{\zeta})+\frac{z(2z-1-\zeta)}{\zeta}\theta(\zeta-z)+\frac{z(2z-1-\bar{\zeta})}{\bar{\zeta}}\theta(\bar{\zeta}-z)
\end{equation}
for the $\mu\leftrightarrow \nu$ symmetric (polarization averaged) part. The corresponding results for 
 the antisymmetric (polarized) part read
\begin{equation}
\tilde F^q= -\frac{N_C\,e_{q}^2}{4\pi^2}(-i \epsilon^{\mu\nu p N})\epsilon^*_\mu (p_{1})\epsilon^*_\nu(p_{2}) 
 \left[\frac{1}{\hat\epsilon}  + \log m^2\right] \tilde F(z,\zeta)\,,
\label{Ftgam}
\end{equation}
with
\begin{equation}
\label{tFz}
 \tilde F(z,\zeta) = \frac{\bar{z}\zeta}{\bar{\zeta}}\theta(z-\zeta)-\frac{\bar{z}\bar{\zeta}}{\zeta}\theta(z-\bar{\zeta})-\frac{z\bar{\zeta}}{\zeta}\theta(\zeta-z)+\frac{z\zeta}{\bar{\zeta}}\theta(\bar{\zeta}-z).
\end{equation}
 Let us stress again that here we concentrate only on the leading logarithmic behaviour and thus
 focus on the divergent parts and their associated logarithmic functions. The ultraviolet divergent parts
 are removed through the renormalization procedure (see for example \cite{Hill})  involving quark and photon correlators 
   ($O^q$, $O^\gamma$) corresponding to one of the two following pairs  
\begin{equation}
\label{OP1}
(\bar q(-\frac{y}{2}N)\gamma.N q(\frac{y}{2}N),\ F^{N\mu}(-\frac{y}{2}N)F_\mu^N(\frac{y}{2}N))\,, 
\end{equation}
or
\begin{equation}
\label{OP2}
(\bar q(-\frac{y}{2}N)\gamma.N\gamma^5 q(\frac{y}{2}N),\ F^{N\mu}(-\frac{y}{2}N)\tilde F_\mu^N(\frac{y}{2}N)).
\end{equation}
The renormalized operators are defined as:
   \begin{eqnarray}
\left(\begin{array}{c} O^q \\ O^\gamma \end{array}\right)_R = \left(\begin{array}{cc} Z_{qq} & Z_{q\gamma} \\ Z_{\gamma q} & Z_{\gamma\gamma} \end{array} \right)\left(\begin{array}{c} O^q \\ O^\gamma \end{array}\right).
\end{eqnarray}
   The matrix element of the renormalized quark-quark correlator is thus equal to
   \begin{eqnarray}
<\gamma(p_1)\gamma(p_2)|O_R^q|0> = Z_{qq}<\gamma(p_1)\gamma(p_2)|O^q|0>+Z_{q\gamma}<\gamma(p_1)\gamma(p_2)|O^\gamma|0>\,,
\end{eqnarray}
with $Z_{qq} = 1+\mathcal{O}\left(\frac{e^2}{\hat{\epsilon}}\right)$. Since the matrix element $<\gamma(p_1)\gamma(p_2)|O^q|0>$ contains a UV divergence (see Eqs.(\ref{Fgam}), (\ref{Ftgam})) and since $<\gamma(p_1)\gamma(p_2)|O^\gamma|0>$ is UV finite and of order $\alpha_{em}^0$, one can absorb this divergence into the renormalization constant $Z_{q\gamma}$. The normalization of the renormalized correlator is fixed  with the help of the renormalization condition which is chosen as
   \begin{eqnarray}
<\gamma(p_1)\gamma(p_2)|O_R^q|0> = 0 \quad {\rm at} \quad M_R=m.
\end{eqnarray}
    In this way the renormalized GDA with vector correlator is equal to
    \begin{equation}
F^q_{R} 
 = -\frac{N_C\,e_{q}^2}{4\pi^2} g_T^{\mu\nu}\epsilon^*_\mu (p_{1})\epsilon^*_\nu(p_{2}) 
  \log{\frac{m^2}{M_{R}^2}} F(z,\zeta) \; 
\label{FRgam}
\end{equation}
and we have a similar result for the renormalized GDA with axial correlator
\begin{equation}
\tilde F^q_{R}
 = -\frac{N_C\,e_{q}^2}{4\pi^2}(-i  \epsilon^{\mu\nu p N})\epsilon^*_\mu (p_{1})\epsilon^*_\nu(p_{2}) 
  \log{\frac{m^2}{M_{R}^2}} \tilde F(z,\zeta) \; .
\label{FRtgam}
\end{equation}
As we want to use QCD factorization formula, which correspond to the factorization scale $M_F$, we identify now the renormalization scale $M_R$ with $M_F$,
\begin{equation}
\label{scale}
M_R = M_F.
\end{equation}
Eqs. (\ref{FRgam}, \ref{FRtgam}) together with  Eqs (\ref{Fz}, \ref{tFz}) permit us to write the expressions of the diphoton generalized distribution amplitudes:
\begin{eqnarray}
\Phi_1^q(z, \zeta, 0) &=& -\frac{N_C\,e_{q}^2}{2\pi^2} \log{\frac{m^2}{M_{F}^2}}F(z, \zeta)\,, \\
\Phi_3^q(z, \zeta, 0) &=& -\frac{N_C\,e_{q}^2}{2\pi^2} \log{\frac{m^2}{M_{F}^2}}\tilde F(z, \zeta).
\end{eqnarray}

Now we are able to write the quark contribution to the $\gamma^* \gamma \to \gamma \gamma$ amplitude at threshold
as a convolution of coefficient functions and GDAs  $\Phi_i^q(z,\zeta,0)$  
\begin{equation}
 W^q_{1}= \int\limits_{0}^1 dz \, C_V^q(z) \, \Phi_1^q(z,\zeta,0)\;\,,~~~~~~~~~~W^q_{3}= \int\limits_{0}^1 dz \,
C_A^q(z)  \,\Phi_3^q(z,\zeta,0)\; ,
\label{fac}
\end{equation}
where the  Born order coefficient functions $C_{V/A}^q$ attached to the quark-antiquark symmetric and 
antisymmetric correlators are equal to:
 \begin{equation}
\label{coeff}
 C_{V}^q = e_q^2\left(\frac{1}{z}-\frac{1}{\bar{z}}\right), \qquad
 C_{A}^q = e_q^2\left(\frac{1}{z}+\frac{1}{\bar{z}} \right)
\end{equation}
We recover in that way the $\ln \frac{m^2}{M_{F}^2}$ term in the right hand side of Eq.(\ref{stupid}).
The photon operator contribution involves  a new coefficient function of order $\alpha_{em}^2$ calculated at the factorization scale $M_{F}$, which
 plays the role of the infrared cutoff. This function is convoluted with the photonic correlators $F^{N\mu}(-\frac{y}{2}N)F_\mu^N(\frac{y}{2}N)$ and $F^{N\mu}(-\frac{y}{2}N)\tilde F_\mu^N(\frac{y}{2}N)$ (cf Eqs.(\ref{Fpa}, \ref{Fpta})) which are of order $\alpha_{em}^0$. These convolutions have the same expressions as those in Section 2 (Eqs.(\ref{W1}, \ref{W3})) with the quark mass replaced by the factorization scale, $m \to M_{F}$. In this way, we recover the second term in the right hand side of Eq.(\ref{stupid}). This equation in fact reflects the independence of the scattering amplitude
 on the choice of the scale $M_{F},$ which is controlled by the renormalization group equation.\\
The factorization scale $ M_{F}^2$ can be chosen in any convenient way. The choice $ M_{F}^2 = Q^2$ kills the logarithmic terms coming from the contribution of photonic GDAs, so that the
scattering amplitude is written (at least in the leading logarithmic approximation) solely in terms of the quark correlators. With $ M_{F}^2 = Q^2$ we interpret this process within the parton model (see e.g.\cite{deWitt}, \cite{FPS} for an analogous interpretation of parton distributions inside the photon).

\section{\normalsize The diphoton GDAs and their QCD evolution equations}
\label{Partevolution}

\begin{figure}[h]
\hspace{3cm}
\includegraphics[width=10cm]{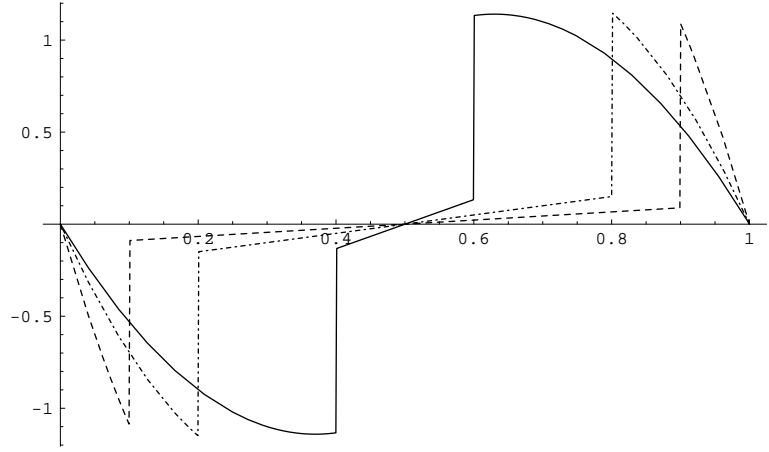}
\caption[]{\small The unpolarized anomalous diphoton GDA $\Phi^q_1 \, 2 \pi^2/(N_C \, e_q^2 \, \log(Q^2/m^2))$ at Born order and at threshold for $\zeta=0.1$ (dashed), $0.2$ (dash-dotted), $0.4$ (solid).}
\label{Figphi1}
\end{figure}

\begin{figure}[h]
\hspace{3cm}
\includegraphics[width=10cm]{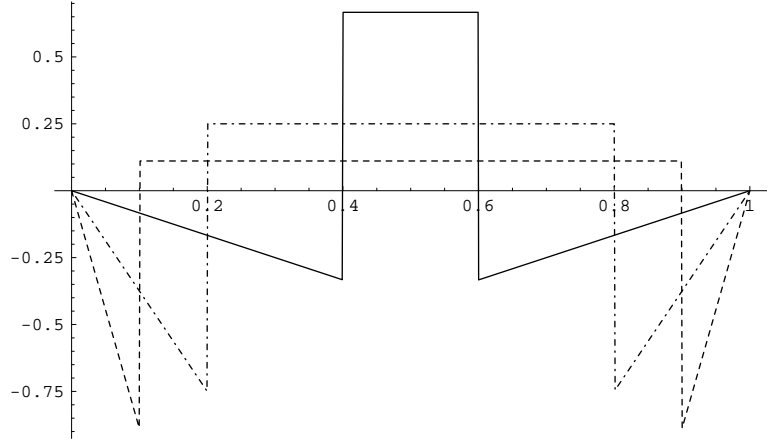}
\caption[]{\small The  polarized anomalous diphoton GDA $\Phi^q_3 \, 2 \pi^2/(N_C \, e_q^2 \, \log(Q^2/m^2))$ at Born order and for $\zeta=0.1$ (dashed), $0.2$ (dash-dotted), $0.4$ (solid).}
\label{Figphi3}
\end{figure}

We have thus demonstrated that it is legitimate to define the  Born order 
diphoton  GDAs at zero $W$ and at $M_F=Q$ as
\begin{eqnarray}
\label{H1}
\Phi_1^q(z,\zeta,0) &=& \frac{N_C\,e_{q}^2}{2\pi^2} \log{\frac{Q^2}{m^2}}\left[\frac{\bar{z}(2z-\zeta)}{\bar{\zeta}}\theta(z-\zeta)+\frac{\bar{z}(2z-\bar{\zeta})}{\zeta}\theta(z-\bar{\zeta}) \right. \nonumber \\
&+&\left.\frac{z(2z-1-\zeta)}{\zeta}\theta(\zeta-z)+\frac{z(2z-1-\bar{\zeta})}{\bar{\zeta}}\theta(\bar{\zeta}-z)\right]
\end{eqnarray}
and
\begin{eqnarray}
\label{H3}
\Phi_3^q(z,\zeta,0) &=& \frac{N_C\,e_{q}^2}{2\pi^2} \log{\frac{Q^2}{m^2}}\left[\frac{\bar{z}\zeta}{\bar{\zeta}}\theta(z-\zeta)-\frac{\bar{z}\bar{\zeta}}{\zeta}\theta(z-\bar{\zeta}) \right. \nonumber \\
&-& \left. \frac{z\bar{\zeta}}{\zeta}\theta(\zeta-z)+\frac{z\zeta}{\bar{\zeta}}\theta(\bar{\zeta}-z)\right].
\end{eqnarray}
Since we focus on the leading logarithmic contribution, we only obtain the {\em anomalous} part of these GDAs. Their $z-$ and $\zeta-$dependence  are shown on Figs.\ref{Figphi1} and \ref{Figphi3}.\\

Note that  $\Phi_i^q(z,\zeta,0)$ GDAs are discontinuous functions of $z$ at 
the points $z= \zeta$ and $z=\bar{\zeta}$. Nevertheless, these GDAs still verify the property of polynomiality :
\begin{eqnarray}
\int_0^1dz\ (2z-1)^n\,\Phi_1^q(z,\ \zeta) &=& \sum_{k=0}^{n+1}a_k\,\zeta^k\ , \\
\int_0^1dz\ (2z-1)^n\,\Phi_3^q(z,\ \zeta) &=& \sum_{k=0}^{n+1}\tilde a_k\,\zeta^k.
\end{eqnarray}
As we have two photons in the final state, the only non vanishing correlations are the C-even i.e. singlet sector of the GDAs which correspond to the combinations of operators $\frac{1}{2}(O^q(x_1,\ x_2)-O^q(x_2,\ x_1))$ and $\frac{1}{2}(\tilde O^q(x_1,\ x_2)+\tilde O^q(x_2,\ x_1))$. This singlet sector means for the GDAs $\Phi_i^q$ the combinations $\Phi_+^q(z,\ \zeta,\ 0) = \frac{1}{2}(\Phi_1^q(z,\ \zeta,\ 0)-\Phi_1^q(\bar{z},\ \zeta,\ 0))$ and $\tilde\Phi_+^q(z,\ \zeta,\ 0) = \frac{1}{2}(\Phi_3^q(z,\ \zeta,\ 0)+\Phi_3^q(\bar{z},\ \zeta,\ 0))$. From the point of view of flavor decomposition, we can distinguish either singlet sector
\begin{equation}
\label{singlet}
\Phi_+^S \propto \sum_{q=1}^{N_f}\Phi_+^q \,,\qquad \tilde\Phi_+^S \propto \sum_{q=1}^{N_f}\tilde\Phi_+^q\,,
\end{equation}
or non-singlet sector
\begin{equation}
\label{nonsinglet}
\Phi_+^{NS} = \Phi_+^q - \Phi_+^{q'} \,,\qquad \tilde\Phi_+^{NS} = \tilde\Phi_+^q - \tilde\Phi_+^{q'}\,,
\end{equation}
where $q$ and $q'$ are two different quark flavors.\\ Let us illustrate the effect of evolution in the simplest, i.e. without mixing with gluons, case of non-singlet and vector GDA $\Phi_+^{NS}$.

Switching on QCD, the non-singlet and vector sector of diphoton GDAs evolves according to the ERBL  evolution equation \cite{BL} modified by the presence of the anomalous part (\ref{H1})
 \begin{equation}
Q^2\,\frac{d}{dQ^2} \Phi_+'^{NS}(z,\zeta, Q^2)= \int_0^1du\ V_{NS}(u, z,
Q^2)\, \Phi_+'^{NS}(u, \zeta, Q^2)\ +\ (e_q^2-e_{q'}^2)\, f'_1(z,\zeta)  \;,
\label{eveq}
\end{equation}
 with $\Phi_+^{NS}(z,\zeta,Q^2)=z(1-z)\Phi_+'^{NS}(z,\zeta,Q^2)$ and
$f_1(z,\zeta)=z(1-z)f'_1(z,\zeta)$.
 Here $f_1(z,\zeta)$ is defined by the r.h.s. of  
 Eqs.(\ref{H1}), as the corresponding functions which 
multiply $e_q^2\ln Q^2/m^2$. The QCD kernel $V_{NS}$ is known \cite{BL} and at the leading order accuracy it reads
\begin{equation}
V_{NS}(u, z, Q^2) = \frac{\alpha_s(Q^2)}{2\pi}C_F 
\left[\frac{u}{z}\theta(z-u)\left(1+\left[\frac{1}{z-u}\right]_+\right) 
+\frac{\bar{u}}{\bar{z}}\theta(u-z)\left(1+\left[\frac{1}{u-z}\right]_+ 
\right)\right]\,,
\label{kernel}
\end{equation} 
with $C_F=(N_C^2-1)/(2N_C)$ and the + prescription is
\begin{equation}
\label{prescription}
\int_0^1da\left[\frac{1}{a}\right]_+f(a) = \int_0^1 da  \, \frac{f(a)-f(0)}{a}.
\end{equation}
The kernel $V_{NS}$ is diagonalized in the basis spanned by Gegenbauer polynomials $C_p^{(3/2)}(x)$ \cite{BKM}. This property permits us to write the solution of the equation (\ref{eveq}) as
\begin{eqnarray}
\label{solution1}
&\phantom{a}&\Phi_+^{NS}(z, \zeta, Q^2) = \\
&\hspace*{-3cm}&z \, (1-z) \sum_{p\ odd}^\infty \left[ A_p\left[\log\frac{Q^2}{m^2}\right]^{-6\frac{\gamma_{qq}(p)}{33-2N_f}} + (e_q^2-e_{q'}^2)\log\frac{Q^2}{m^2}\frac{f'_p(\zeta)}{1+6\frac{\gamma_{qq}(p)}{33-2N_f}}\right] C_p^{(3/2)}(2z-1), \nonumber
\end{eqnarray}
where $A_p$ are integration constants, $N_f$ is the number of flavors, $\gamma_{qq}(p)$ are the usual anomalous dimensions
\begin{equation}
\label{dimano}
\gamma_{qq}(p) = C_F\left(\frac{1}{2}-\frac{1}{(p+1)(p+2)}+2\sum_{k=2}^{p+1}\frac{1}{k}\right)
\end{equation}
and the coefficients $f'_p(\zeta)$ are projection of $f'_1(z, \zeta)$ on appropriate $C_p^{(3/2)}$ polynomials
\begin{equation}
\label{coef}
f'_p(\zeta) = \frac{4(2p+3)}{(p+1)(p+2)}\int_0^1dz\ z \, \bar{z} \, f'_1(z, \zeta)C_p^{(3/2)}(2z-1).
\end{equation}
Since the expressions for $f'_p(\zeta)$ are lenghty, we do not presente them explicitely. For large $Q^2$, the solution (\ref{solution1}), at the leading logarithm level, reads
\begin{equation}
\label{solapprox}
\Phi_+^{NS}(z, \zeta, Q^2) \simeq (e_q^2-e_{q'}^2)\log\frac{Q^2}{m^2} \, z \,\bar{z}\sum_{p\ odd}^\infty\frac{f'_p(\zeta)}{1+6\frac{\gamma_{qq}(p)}{33-2N_f}}C_p^{(3/2)}(2z-1).
\end{equation}
The formula (\ref{solapprox}) shows the known result that the anomalous part of GDAs dominates at large $Q^2$. The $ z, \zeta$ dependences of $\Phi_1^q$ are nevertheless modified by strong interaction, due to the presence of the denominator $1+6\frac{\gamma_{qq}(p)}{33-2N_f}$.

\begin{figure}[!h]
\centering
\includegraphics[width=10cm]{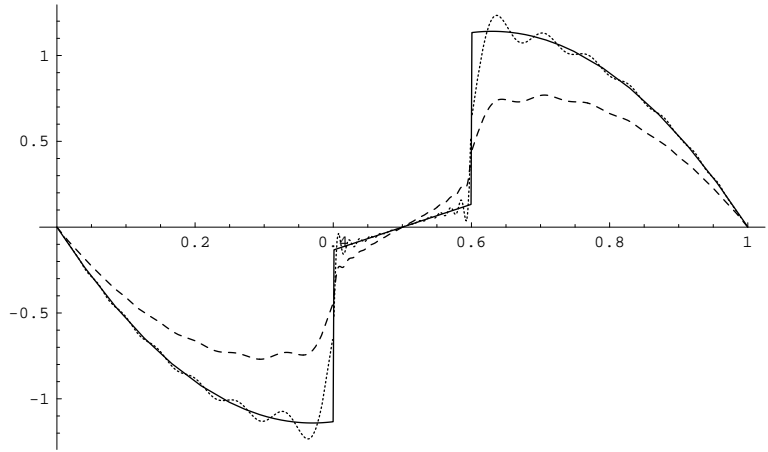}
\caption{\small The bare (solid line)
and the QCD evolved (dashed line) non-singlet and vector sector of diphoton GDA  at large $Q^2$   $\Phi^{NS}_+/((e_q'^2-e_q^2) \, \log(Q^2/m^2))$ (see Eq.(\ref{solapprox})) for $\zeta=0.2$. The dotted line is the truncated (see text) Gegenbauer expansion of the bare GDA.}
\label{FigsolGDA}
\end{figure}

The result of this method, for $N_f=2$, is shown in Fig.\ref{FigsolGDA}. The procedure
 of decomposing a GDA into series of Gegenbauer polynomials ensures a good numerical description of both the unevolved GDA
and the one with the evolution taken into account, except in the regions
$z$ close to $\zeta$ and $1-\zeta$ (it is well known that theta function are very badly described by series of polynomials). Fig.\ref{FigsolGDA} was obtained after combining the
sum of 41 (201) contributions of Gegenbauer polynomials for the regions $0<z<\zeta$ and $1-\zeta<z<1$ ($\zeta<z<1-\zeta$). Such
a choice for the regions $0<z<\zeta$ and $1-\zeta<z<1$ is motivated by numerical instabilities of the Gegenbauer series
(\ref{solapprox}) when $z$ is close to $0$ and $1,$ which requires to truncate the series at a moderate number of terms (41 terms), which nevertheless leads to a good stability of results. These numerical instabilities are due to the fact that, as soon as $z$ deviates from 0 or 1, $f'_p(\zeta)$ (Eq.(\ref{coef})) becomes
very large with $\zeta$ fixed and $p$ larger than 50. This is in contrast with
the region $\zeta<z<1-\zeta$ where a very good stability of the results
is achieved when summing a very high number of terms (we took 201 terms: taking more terms would then generate huge instablities in the vicinity of $\zeta$ and $1-\zeta$).

As can be seen from Fig.\ref{FigsolGDA},  QCD evolution affects the form of the GDA in the whole
$z$ range.

\begin{figure}[h]
\hspace{3cm}
\includegraphics[width=10cm]{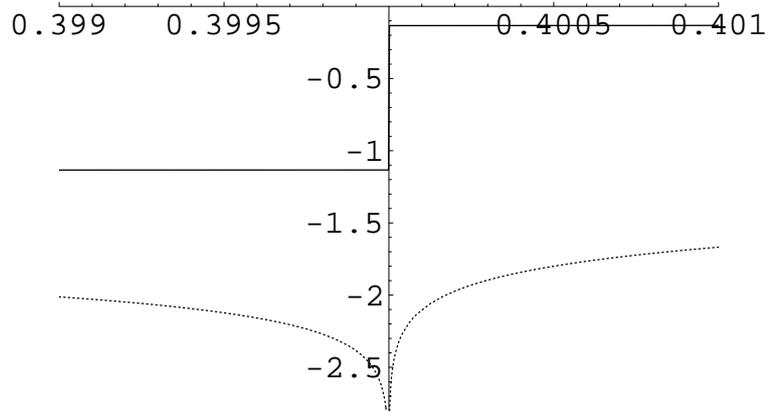}
\caption[]{\small The behaviour of the evolved GDA from the leading logarithmic resummations (\ref{solutionphi1A}, \ref{solutionphi2A}) in the vicinity of the discontinuity point  $z=\zeta,$ for the case $\zeta=.4$ (dotted curve). The Born GDA is given for reference as the solid line.}
\label{asympEvolution}
\end{figure}

In practice, one cannot trust the truncated Gegenbauer expansion near the discontinuity points.  To obtain a trustable behaviour of the evolved GDA around $z=\zeta$ and $z=\bar{\zeta},$ one has to
sum in Eq.(\ref{solapprox}) the infinite series involving Gegenbauer  polynomials, which is a non trivial task.
Because of that, one could think about a method based on direct iteration of the non-homogeneous equation (\ref{eveq}), keeping only the leading terms at large $Q^2$ and we refer for all details to the Appendix. In particular, we have solved the evolution equation (\ref{eveq}, \ref{kernel}) in the vicinity of singularity points $z=\zeta, \, 1-\zeta.$  Because of the antisymmetry of the GDA (\ref{H1}), let us concentrate our discussion on the form of the evolved GDA around the singular point $z=\zeta.$
The iteration of the kernel  (\ref{eveq}) generates leading terms of the form
$K^n \ln^n(\zeta-z)$ for $z \to \zeta^-$ and $K^n \ln^n(z-\zeta)$ for $z \to \zeta^+\,,$
which we resummed\footnote{Note that this series corresponds to the resummation of soft gluons exchanged between quark-antiquark lines, when iterating the ERBL kernel on the bare GDA.} in
\beq
\label{solutionphi1A}
\Phi^{NS}_+(z \to \zeta^-,\zeta)=\frac{1}{2} \, (e_q^2-e_{q'}^2) \, \log \frac{Q^2}{m^2}
\left[
\frac{4 \zeta^2-\zeta-1}{\bar{\zeta}} (1-K \ln(\zeta-z)) +
\frac{1}{1-K \ln(\zeta-z)}\right] 
\eq
and 
\beq
\label{solutionphi2A}
\Phi^{NS}_+(z \to \zeta^+,\zeta)=
\frac{1}{2}\,(e_q^2-e_{q'}^2) \, \log \frac{Q^2}{m^2}
\left[
\frac{4 \zeta^2-\zeta-1}{\bar{\zeta}} (1-K \ln(z-\zeta)) -
\frac{1}{1-K \ln(z-\zeta)}\right]\,,
\eq
where
$K=6\, C_F/(11 N_C-2 N_f)\,.$
We display in Fig.\ref{asympEvolution} the asymptotic resummation given by Eqs.(\ref{solutionphi1A},\ref{solutionphi2A}).
As mentionned above, the resummation of logarithmic contribution is required in the regions $K \, \log |z-\zeta| \gtrsim 1$
and $K \, \log |z-\bar{\zeta}|\gtrsim 1,$ which in our case ($K=8/29$) corresponds to $|z-\zeta|, |z-\bar{\zeta}| \lesssim 3. \ 10^{-2}.$
In practice, subleading contributions are non negligible in a larger domain
in $z$ around $\zeta$ and $\bar{\zeta},$ and prevent us from getting trustable numerical results based on this iterative method.
This is due to the numerically large values of $K.$ This iterative method stabilizes only when applied for unphysically small values of $K$ (typically $\lesssim .03$).

Note that these singularities of the GDA at $z=\zeta$ and $z=1-\zeta$ does not affect
the DVCS amplitude (\ref{fac}) since they are integrable singularities\footnote{The coefficient function
 $C_{V}^q$ (\ref{coeff}) is regular. This would remain true after taking into account its QCD evolution}.

\section{\normalsize  Conclusions}

We derived the leading amplitude of the DVCS (polarization averaged or polarized) process on
a photon target at threshold. We have shown that the amplitude coefficients $W_i^q$ factorize in the forms shown in
Eq.(\ref{fac}),
irrespectively of the fact that the handbag diagram interpretation appears only 
{\em after} cancellation of UV divergencies in the scattering amplitude.
We have shown that the objects $\Phi_i^q(z,\zeta,0)$ are matrix elements of non-local quark operators 
on the light cone, and that they have an anomalous component which is proportional to $\log(Q^2/m^2)$.
They thus have all the properties attached to generalized  distribution amplitudes, and 
they obey {\em non-homogeneous} ERBL evolution equations. This new type of evolution equations is an interesting playground to study the effects of gluon radiation on a non diagonal object such as a GDA.

\section*{Acknowledgments}

\noi
We are grateful to Igor Anikin, Markus Diehl, Samuel Friot and Jean Philippe Lansberg for useful 
discussions and correspondance. 
This work is partly supported by the French-Polish scientific agreement Polonium 7294/R08/R09,  the ECO-NET program, contract 
18853PJ, and the ANR-06-JCJC-0084-02.

\section*{Appendix}

Using
\beq
\label{alphas}
\alpha_s(Q^2)=\frac{12 \pi}{(11 N_C-2 N_f) \ln \frac{Q^2}{\Lambda^2}}\,,
\eq
denoting
$t=\log\frac{Q^2}{m^2}$ and redefining the GDA as
$\Phi_q(z,\zeta,t)=z\, \bar{z}\, \Phi_q'(z,\zeta,t),$
the evolution equation (\ref{eveq}) reads
\begin{equation}
\label{evolutionfp}
\frac{\partial}{\partial t}\Phi_+'^{NS}(z,\zeta,t) = \int_0^1du \, V_{NS}(u,z)\, \Phi_+'^{NS}(u,\zeta,t) + (e_q^2-e_{q'}^2)f'_1(z,\zeta)\,,
\end{equation}
where $f'_1(z,\zeta),$ with the help of Eq.(\ref{H1}), is
\begin{eqnarray}
\label{deff1p}
f'_1(z,\zeta) &=& \frac{N_C}{2\pi^2} \left[\frac{(2z-\zeta)}{z\bar{\zeta}}\theta(z-\zeta)+\frac{(2z-\bar{\zeta})}{z\zeta}\theta(z-\bar{\zeta}) \right. \nonumber \\
&+&\left.\frac{(2z-1-\zeta)}{\bar z\zeta}\theta(\zeta-z)+\frac{(2z-1-\bar{\zeta})}{\bar z\bar{\zeta}}\theta(\bar{\zeta}-z)\right]\,.
\end{eqnarray}
The solution of Eq.(\ref{evolutionfp}) is the sum
\begin{equation}
\Phi_+'^{NS}(z,\zeta,t) = g(z,\zeta,t)+i(z,\zeta,t)\,,
\end{equation}
where $g(z,\zeta,t)$ is the general solution of the homogeneous evolution equation,
\begin{equation}
\label{eqhomo}
\frac{\partial}{\partial t}g(z,\zeta,t) = \int_0^1du \, V_{NS}(u,z,t)\,g(u,\zeta,t)
\end{equation}
and $i(z,\zeta,t)=t\ h(z,\zeta)$ is a particular solution of the non-homogeneous evolution equation
\begin{equation}
\label{eqh}
h(z,\zeta) = t\int_0^1du\, V_{NS}(u,z) \, h(u,\zeta)+h^{(0)}(z,\zeta)\,,
\end{equation}
where $h^{(0)}(z,\zeta)= (e_q^2-e_{q'}^2)f'_1(z,\zeta) \,.$
As we will be interested in the leading terms for large $Q^2,$ the general solution of the homogeneous
equation (\ref{eqhomo}) can be omitted.
Eq.(\ref{eqh}) reads
\begin{equation}
\label{eqhK}
h(z,\zeta) = K \int_0^1du\, \bar{V}(u,z) \, h(u,\zeta)+h^{(0)}(z,\zeta)\,,
\end{equation}
with 
$K=6\, C_F/(11 N_C-2 N_f)$ and $\bar{V}$ is the expression inside $[ \cdots]$ in Eq.(\ref{kernel}). This equation may be solved numerically by the method of successive iterations,
which represents the solution as
\beq
\label{devsolution}
h(z,\zeta) =  \sum_{n=0}^{\infty}(K \bar{V})^n \otimes h^{(0)}\,.
\eq

Since we discuss the vector GDA, satisfying Eq.(\ref{evolutionfp}), which is antisymmetric with respect
to the substitution $z \to \zb,$ it is sufficient to construct the solution for $z \in [0,1/2]\,.$ Similarly, the solution is symmetric with respect
to the substitution $\zeta \to \bar{\zeta},$ and thus without loss of 
generality, we restrict our analysis to $\zeta \in [0,1/2]\,.$

We write the solution as
\beq
\label{secteursolution}
h(z,\zeta) = h_1(z,\zeta) \, \theta(\zeta-z)+ h_2(z,\zeta)\, \theta(z-\zeta)\,,
\eq
and use the same decomposition for $h^{(0)}(z,\zeta)$
\beq
\label{secteursolutionh0}
h^{(0)}(z,\zeta) = h^{(0)}_1(z,\zeta) \, \theta(\zeta-z)+ h^{(0)}_2(z,\zeta)\, \theta(z-\zeta)\,,
\eq
where, with the help of Eq.(\ref{deff1p}),
\beq
\label{secteursolutionh120}
 h^{(0)}_1(z,\zeta) =  (e_q^2-e_{q'}^2) \, \frac{1}{\bar{u}} \left( \frac{2 u -1}{\zeta \bar{\zeta}} -2 \right) \quad {\rm and }  \quad h^{(0)}_2(z,\zeta)=(e_q^2-e_{q'}^2) \,\frac{\zeta}{\bar{\zeta}}\frac{(2 u-1)}{u \, \bar{u}} \,.
\eq
In these notations, Eq.(\ref{eqhK}) reads
\bea
\label{eqhsecteur1}
h_1(z,\zeta)&=& K \left[  \int_0^z \left\{\left[ \frac{1-2 z}{z \, \bar{z}} u - \frac{u}{\bar{z}} \frac{1}{\bar{z} -u} \right] \, h_1(u,\zeta) + \frac{\frac{u}{z} h_1(u,\zeta) - h_1(z,\zeta)}{z-u}   \right\} du \right.  \\
&+& \left. \int_z^\zeta \left\{\left[ \frac{1-2 u}{\bar{z}}  - \frac{u}{\bar{z}} \frac{1}{\bar{z} -u} \right] \, h_1(u,\zeta) + \frac{\frac{\bar{u}}{\bar{z}} h_1(u,\zeta) - h_1(z,\zeta)}{u-z}   \right\} du \right. \nonumber \\
&+& \left. \frac{1}{\bar{z}}\int_\zeta^{1/2} \left[ 1-2 u + \frac{\bar{u}}{u-z}- \frac{u}{\bar{z}-u}\right] \, h_2(u,\zeta)  du \right] +h_1^{(0)}(z,\zeta)\,,\nonumber \\
\label{eqhsecteur2}
h_2(z,\zeta)&=& K \left[  \int_0^\zeta \left[ \frac{1-2 z}{z \, \bar{z}} u + u\left( \frac{1}{z(z -u)}- \frac{1}{\bar{z}(\bar{z} -u)} \right)\right] \, h_1(u,\zeta)  du \right.  \\
&+& \left. \int_\zeta^z \left\{\frac{1-2 z}{z \bar{z}} u   \, h_2(u,\zeta) +\frac{\frac{u}{z} h_2(u,\zeta)-h_2(z,\zeta)}{z-u}-\frac{\frac{u}{\bar{z}} h_2(u,\zeta)+h_2(z,\zeta)}{\bar{z}-u}
\right\} du \right. \nonumber \\
&+& \left. \int_z^{1/2} \left[ \frac{1-2 u}{\bar{z}}  h_2(u,\zeta) -\frac{\frac{u}{\bar{z}} h_2(u,\zeta)+h_2(z,\zeta)}{\bar{z}-u}+\frac{\frac{\bar{u}}{\bar{z}} h_2(u,\zeta)-h_2(z,\zeta)}{u-z}
  \right] du \right] +h_2^{(0)}(z,\zeta)\,.\nonumber
\eea
In order to understand the nature of the singularity observed at $z=\zeta$ and $z=\bar{\zeta},$
it is useful to observe that the dominant contribution appearing at each iteration involves
$K^n \ln^n(\zeta-z)$ for $z \to \zeta^-$ and $K^n \ln^n(z-\zeta)$ for $z \to \zeta^+\,.$
Our aim is now to sum all such leading logarithmic terms. For that, we note
that such terms are generated by the approximate evolution equations, obtained from (\ref{eqhsecteur1})
and (\ref{eqhsecteur2}) when keeping only terms generating large $\ln^n|z-\zeta|\,:$
\beqa
\label{eqhsecteur1A}
\tilde{h}_1(z,\zeta)&=& K \left[  \int_0^z  \frac{ \tilde{h}_1(u,\zeta) - \tilde{h}_1(z,\zeta)}{z-u}   du 
+  \int_\zeta^{1/2} \frac{\tilde{h}_2(u,\zeta)}{u-z}  du \right] +h^-_0\,,
\\
\label{eqhsecteur2A}
\tilde{h}_2(z,\zeta)&=& K \left[  \int_0^\zeta \frac{\tilde{h}_1(u,\zeta)}{z -u}+ 
 \int_z^{1/2} \frac{ \tilde{h}_2(u,\zeta)-\tilde{h}_2(z,\zeta)}{u-z}
   du \right] +h^+_0\,,
\eqa
where $h^+_0= (2 \zeta-1)/\bar{\zeta}^2$ and $h^-_0= h^+_0-1/(\zeta \bar{\zeta})\,,$ see Eq.(\ref{secteursolutionh120}).
The solution of these equations are
\beqa
\label{solution1A}
\tilde{h}_1(z,\zeta)&=&\frac{1}{2}\left((h_0^-+h_0^+)(1-K \ln(\zeta-z)) + \frac{h_0^--h_0^+}{1-K \ln(\zeta-z)}\right) \,,\\
\label{solution2A}
\tilde{h}_2(z,\zeta)&=&\frac{1}{2}\left((h_0^-+h_0^+)(1-K \ln(z-\zeta)) - \frac{h_0^--h_0^+}{1-K \ln(z-\zeta)}\right)\,.
\eqa
Identifying
\beqa
\label{relphi1h1}
\Phi^{NS}_+(z \to \zeta^-,\zeta)&=&z \, \bar{z} \,(e_q^2-e_{q'}^2) \, \log \frac{Q^2}{m^2} \, \tilde{h}_1(z,\zeta)\,,\\
\label{relphi2h2}
\Phi^{NS}_+(z \to \zeta^+,\zeta)&=&z \, \bar{z} \,(e_q^2-e_{q'}^2) \, \log \frac{Q^2}{m^2} \, \tilde{h}_2(z,\zeta)
\eqa
we recover Eqs.(\ref{solutionphi1A}, \ref{solutionphi2A}) of section \ref{Partevolution}.

Before proving that (\ref{solution1A}, \ref{solution2A}) are required solutions, let us explain first,  in a  somehow heuristic way, how one can guess the form of solutions (\ref{solution1A}, \ref{solution2A}) from Eq.(\ref{eqhsecteur1A}) and (\ref{eqhsecteur2A}).
We rewrite the second (resp. first) integral of Eq.(\ref{eqhsecteur1A}) (resp. (\ref{eqhsecteur2A})) using $\tilde{h}_{2(1)}(u,\zeta)=\tilde{h}_{2(1)}(u,\zeta)-\tilde{h}_{2(1)}(z,\zeta)+\tilde{h}_{2(1)}(z,\zeta)$ and perform the integration of the constant term at leading logarithmic order, which gives
\beqa
\label{eqhsecteur1Astep1}
\hspace{-.6cm}\tilde{h}_1(z,\zeta)\!\!\!\!\!&=& \!\!\!\!\!\!K \!\!\left[  \int_0^z  \frac{ \tilde{h}_1(u,\zeta) - \tilde{h}_1(z,\zeta)}{z-u}   du 
+  \!\int_\zeta^{1/2} \frac{\tilde{h}_2(u,\zeta)-\tilde{h}_2(z,\zeta)}{u-z}  du \right]\!\!\! -K \, \tilde{h}_2(z,\zeta) \, \ln(\zeta -z) +h^-_0\,, \\
\label{eqhsecteur2Astep1}
\hspace{-.6cm}\tilde{h}_2(z,\zeta)\!\!\!\!\!&=&\!\!\!\!\!\! K \!\!\left[  \int_0^\zeta \frac{\tilde{h}_1(u,\zeta)-\tilde{h}_1(z,\zeta)}{z -u}du+ 
 \!\int_z^{1/2} \frac{ \tilde{h}_2(u,\zeta)-\tilde{h}_2(z,\zeta)}{u-z}
   du \right]\! \!\!-K \, \tilde{h}_1(z,\zeta) \, \ln(\zeta -z)+h^+_0\,.
\eqa
The trick, valid at leading logarithmic order, is now to replace in the denominator of the integrand of the right hand side of these equations $z$ by $\zeta$ and to replace the lower 
(resp. upper) bound of the 
second (resp. first) integral of  Eq.(\ref{eqhsecteur1Astep1}) (resp. (\ref{eqhsecteur2Astep1})) by $z,$ which gives
\beqa
\label{eqhsecteur1Astep2}
\hspace{-.6cm}\tilde{h}_1(z,\zeta)\!\!\!\!\!&=& \!\!\!\!\!K \!\!\left[  \int_0^z  \frac{ \tilde{h}_1(u,\zeta) - \tilde{h}_1(z,\zeta)}{\zeta-u}   du 
+  \!\int_z^{1/2} \frac{\tilde{h}_2(u,\zeta)-\tilde{h}_2(z,\zeta)}{u-\zeta}  du \right]\!\! -\!K \, \tilde{h}_2(z,\zeta) \, \ln(\zeta -z) +h^-_0 \,,\\
\label{eqhsecteur2Astep2}
\hspace{-.6cm}\tilde{h}_2(z,\zeta)\!\!\!\!\!&=& \!\!\!\!\!K \!\!\left[  \int_0^z \frac{\tilde{h}_1(u,\zeta)-\tilde{h}_1(z,\zeta)}{\zeta -u}du+ \!
 \int_z^{1/2} \frac{ \tilde{h}_2(u,\zeta)-\tilde{h}_2(z,\zeta)}{u-\zeta}
   du \right] \!\!-\!K \, \tilde{h}_1(z,\zeta) \, \ln(z-\zeta)+h^+_0\,.
\eqa
Writing symbolically 
\beq
\label{defSigmaDelta}
\Sigma=\tilde{h}_1+\tilde{h}_2 \quad {\rm and } \quad \Delta=\tilde{h}_1-\tilde{h}_2,
\eq
these two equations decouple in leading logarithmic accuracy into 
\beqa
\label{eqDelta}
\Delta(z,\zeta)&=&  K \, \Delta(z,\zeta)\, \ln(z-\zeta)
 +h^-_0 -h^+_0 \,,\\
\label{eqSigma}
\Sigma(z,\zeta)&=& 2 \, K  \int_0^z  \frac{ \Sigma(u,\zeta) - \Sigma(z,\zeta)}{\zeta-u}   du 
 -K \, \ln(z-\zeta)+h^-_0 +h^+_0 \,.
\eqa
The algebraic equation (\ref{eqDelta}) is solved in
\beq
\label{solDelta}
\Delta(z,\zeta)= \frac{h^-_0 -h^+_0}{1- K \, \ln(z-\zeta)}\,.
\eq
Eq.(\ref{eqSigma}) can be turned into an elementary differential equation when integrating 
the second term of the integral and differentiating with respect to $z,$ giving
\beq
\label{EqSigma1}
\Sigma' [1-K \, \ln(\zeta-z)]+ \Sigma(z) \, \frac{K}{\zeta-z}=2 \, K \, \frac{\Sigma}{\zeta-z}\,,
\eq
which solution is
\beq
\label{solSigma}
\Sigma(z,\zeta)= C \, [1- K \, \ln(\zeta-z)]\,.
\eq
From Eq.(\ref{defSigmaDelta}),
\beq
\label{defSigmaDeltaInverse}
\tilde{h}_1=\frac{\Sigma+\Delta}{2}  \quad {\rm and } \quad \tilde{h}_2=\frac{\Sigma-\Delta}{2}\,,
\eq
and the constant $C$ is fixed as $C=h^-_0 +h^+_0$ by comparison of the lowest order term in the $K$ expansion
with Eqs.(\ref{eqhsecteur1A}, \ref{eqhsecteur1A}).
This justifies the ansatz (\ref{solution1A}, \ref{solution2A}).

We now prove 
 that the expressions (\ref{solution1A}, \ref{solution2A}) 
are the wanted solutions, by direct substitution of them into 
 the equations (\ref{eqhsecteur1A}, \ref{eqhsecteur2A}) and  performing necessary integrations. 
Let us consider as example the Eq.(\ref{eqhsecteur1A})   for $\tilde{h}_1(z,\zeta)$. The right 
hand side of it takes thus the form 
\begin{eqnarray}
\label{rhsh1}
&&\hspace*{-0.5cm}\mbox{rhs of Eq.~(\ref{eqhsecteur1A})}=\frac{K^2}{2}\int\limits_0^z\,\frac{du}{z-u}\left[
-(h_0^-+h_0^+)\,\ln \frac{\zeta-u}{\zeta-z} + \frac{(h_0^--h_0^+)\,
\ln \frac{\zeta-u}{\zeta-z}}{[1-K \ln (\zeta-u)][1-K\ln (\zeta -z)]}
\right]\nonumber \\
&& + \frac{K}{2}\int\limits_\zeta^{1/2}\,\frac{du}{u-z}\left[
(h_0^-+h_0^+)(1-K\ln (u-\zeta))-\frac{(h_0^--h_0^+)}{1-K \ln (u-\zeta)}
\right]+h_0^-\;.
\end{eqnarray}
The three integrals which appear in (\ref{rhsh1}) are calculated in a straightforward way 
\beq
\label{Ia}
\int\limits_0^z\,\frac{du}{z-u}\,\ln\frac{\zeta-u}{\zeta-z} = -Li_2(1-\frac{\zeta}{\zeta-z})
\underset{\zeta-u\to 0}{\longrightarrow}
 \frac{1}{2}\ln^2 (\zeta -z)\,,
\eq
\beq
\label{Ic1}
\int\limits_\zeta^{1/2}\,\frac{du}{u-z}=\ln \frac{1/2-z}{\zeta-z} 
\underset{\zeta-u\to 0}{\longrightarrow}
 -\ln (\zeta - z)\,,
\eq
\beq
\label{Ic2}
\int\limits_\zeta^{1/2}\,\frac{du}{u-z}\,\ln (u-\zeta)=
\ln(1/2-\zeta)\ln(1+ \frac{1/2-\zeta}{\zeta -z})+Li_2\left(-\frac{1/2-\zeta}{\zeta-z}  \right)
\underset{\zeta-u\to 0}{\longrightarrow}
-1/2\ln^2 (\zeta -z)\,.
\eq

The integral $\int\limits_\zeta^{1/2}\,\frac{du}{u-z}\,\frac{1}{1-K\ln (u-\zeta)}$ can be evaluated by 
developing $\frac{1}{1-K\ln (u-\zeta)}$ in its Taylor series in $K\ln (u-\zeta)$ and
 by performing integration 
over $u$ in each term of this expansion. The leading terms in
 the limit of small values of $\zeta -z$ are then summed up again, which leads to the formula 
\beq
\label{Id}
\int\limits_\zeta^{1/2}\,\frac{du}{u-z}\,\frac{1}{1-K\ln (u-\zeta)}
\underset{\zeta-u\to 0}{\longrightarrow}
 \frac{1}{K}\ln \left[
1-K \ln (\zeta -z)
 \right]\;.
\eq
More involved is the evaluation of the integral
\[
I=\int\limits_0^z\,\frac{du}{z-u}\,\frac{\ln \frac{\zeta-u}{\zeta-z}}{1-K\ln(\zeta-u)}\;.
\]
We calculate it by performing first the change of variables $\tau = \frac{\zeta-z}{\zeta-u}$ which 
leads to the expression
\beq
\label{Ib}
I=\frac{G}{K}\int\limits_{\frac{1}{\zeta}(\zeta-z)}^1\,\frac{d\tau}{\tau(1-\tau)}\,
\frac{\ln \tau}{1+G \ln \tau}\;, \;\;\;\;\;\;\mbox{with}\;\;\;\;\;\;G=\frac{K}{1-K\ln (\zeta - z)}\,.
\eq
Next, we represent the integrand of $I$ as a double series in $\tau$ and $G\ln \tau$ and perform 
the integration over $\tau$ in each term of this double series by keeping leading terms in the limit
$\zeta -z \to 0\,:$
\begin{eqnarray}  
\label{Ibfinal}
I&=&\frac{G}{K}\sum\limits_{n=0}^\infty (-1)^{n+1}G^n \sum\limits_{p=0}^\infty
\int\limits_{\frac{1}{\zeta}(\zeta-z)}^1 d\tau\, \tau^{p-1}\ln^{n+1}\tau 
\underset{\zeta-u\to 0}{\longrightarrow} \frac{1}{GK}\sum\limits_{n=0}^\infty (-1)^{n+2}G^{n+2}
\frac{\ln^{n+2}(\zeta -z)}{n+2} 
\nonumber \\
&=& \frac{1}{K^2}\left(1-K\ln (\zeta-z) \right)\left[
\ln(1-K\ln(\zeta-z)) +\frac{K\ln (\zeta-z)}{1-K\ln(\zeta-z)}
    \right]\;.
\end{eqnarray}
By substituting expressions (\ref{Ia}, \ref{Ic1}, \ref{Ic2}, \ref{Id}, \ref{Ibfinal}) into 
Eq.(\ref{rhsh1}) we recover the expression (\ref{solution1A})
for $\tilde{h}_1(z,\zeta)$. In a similar way we have checked that the expressions (\ref{solution1A})
and (\ref{solution2A}) satisfy the second approximated evolution equation 
(\ref{eqhsecteur2A})
for $\tilde{h}_2(z,\zeta)$.


\end{document}